\documentclass{article}
\usepackage{graphicx}
\usepackage{amsmath}

\PassOptionsToPackage{numbers, compress}{natbib}

\usepackage[preprint]{neurips_2024}

\usepackage[utf8]{inputenc} 
\usepackage[T1]{fontenc}    
\usepackage{hyperref}       
\usepackage{url}            
\usepackage{booktabs}       
\usepackage{amsfonts}       
\usepackage{nicefrac}       
\usepackage{microtype}      
\usepackage{xcolor}         
\usepackage{amsmath}
\usepackage{bbm}
\usepackage{amssymb}
\usepackage{enumitem}
\usepackage{caption}
\usepackage{multirow}

\title{PharMolixFM: All-Atom Foundation Models\\ for Molecular Modeling and Generation}

\author{%
    \textbf{Yizhen Luo}$^{1,2,}$\thanks{Equal Contribution.}\ \,$^{,}$\thanks{Work done during internship in PharMolix Inc.}\ \,, 
    \textbf{Jiashuo Wang}$^{1,2,}$\footnotemark[1]\enspace, 
    \textbf{Siqi Fan}$^{2,1}$, 
    \textbf{Zaiqing Nie}$^{2,1,}$\thanks{Corresponding author}\\
    $^{1}$PharMolix Inc. \\
    $^{2}$Institute for AI Industry Research (AIR), Tsinghua University\\
    \texttt{\{yz-luo22, wang-js21\}@mails.tsinghua.edu.cn}\\
    \texttt{\{fansiqi, zaiqing\}@air.tsinghua.edu.cn}\\
}

\begin{document}

\maketitle

\begin{abstract}
Structural biology relies on accurate three-dimensional biomolecular structures to advance our understanding of biological functions, disease mechanisms, and therapeutics. While recent advances in deep learning have enabled the development of all-atom foundation models for molecular modeling and generation, existing approaches face challenges in generalization due to the multi-modal nature of atomic data and the lack of comprehensive analysis of training and sampling strategies. To address these limitations, we propose PharMolixFM, a unified framework for constructing all-atom foundation models based on multi-modal generative techniques. Our framework includes three variants using state-of-the-art multi-modal generative models. By formulating molecular tasks as a generalized denoising process with task-specific priors, PharMolixFM achieves robust performance across various structural biology applications. Experimental results demonstrate that PharMolixFM-Diff achieves competitive prediction accuracy in protein-small-molecule docking (83.9\% vs. 90.2\% RMSD < 2Å, given pocket) with significantly improved inference speed. Moreover, we explore the empirical inference scaling law by introducing more sampling repeats or steps. Our code and model are available at \url{https://github.com/PharMolix/OpenBioMed}.
\end{abstract}

\section{Introduction}

Structural biology aims to analyze the three-dimensional structures and dynamics of biomolecules, including small molecules, nucleic acids, and proteins \cite{congreve2005keynote}. Accurate molecular structures enable insights into various applications such as understanding protein functions and biological processes \cite{whisstock2003prediction}, unveiling the mechanisms of human diseases \cite{yan2005mechanisms}, and identifying potential therapeutic agents \cite{surade2012structural}. Due to the high cost of obtaining and analyzing biomolecular structures in virto, machine learning approaches \cite{jumper2021highly, baek2021accurate, dauparas2022robust} have been widely adopted in this field. Thanks to the growing biomolecular structure databases \cite{burley2017protein, varadi2022alphafold,lin2023evolutionary}, deep learning models have demonstrated great success in modeling and generating accurate molecular structures. 

Innovated by recent advancements of foundation models in computer vision (CV) \cite{awais2025foundation, yuan2021florence} and natural language processing (NLP) \cite{zhou2024comprehensive, touvron2023llama}, the development of molecular foundation models capable of solving various tasks in structural biology has attracted rising research interest. Most research works focus on unifying different types of biomolecules by capturing the fundamental principles of biomolecular structures such as the chemical bonds and van der Waals forces at the atom level. For example, AlphaFold3 \cite{abramson2024accurate} decomposes molecules, proteins, RNAs, and DNAs as atoms and trains a denoising diffusion probabilistic model \cite{ho2020denoising} for accurate prediction of monomer and complex structures. RossetaFold-AA \cite{rfaa} develops a three-track architecture to fuse 1D, 2D, and 3D information and predict all-atom coordinates. ESM-AA \cite{zheng2024esm}, on the other hand, proposes to train a language model bridging proteins and small molecules by randomly unzipping part of the protein sequence. 

Despite their advantages in capturing molecular structures and interactions, existing models still fall short in their generalization capability across molecular modeling and generation tasks due to the following challenges. (1) First, differing from text and image data that exhibit a single modality, each atom comprises both a discrete atom type and multiple continuous coordinates. This poses challenges for structure models to effectively integrate and predict the multi-modal information of atoms. (2) Besides, there lacks an in-depth analysis of how different training objectives and sampling strategies may impact the performance of all-atom foundation models.   

To address these challenges and further expand the application scope of deep learning models in structural biology, we present \textbf{PharMolixFM}, a unified framework for developing all-atom foundation models for molecular modeling and generation. We implement state-of-the-art multi-modal generative models, including multi-modal diffusion \cite{austin2021structured} (PharMolixFM-Diffusion), multi-modal flow matching \cite{campbell2024generative} (PharMolixFM-Flow), and Bayesian flow networks \cite{graves2023bayesian} (PharMolixFM-BFN), to jointly capture atom types and atom coordinates within different types of biomolecules. Innovated by PocketXMol \cite{pocketxmol}, we envision that different downstream tasks can be unified by a generative denoising process with different priors. This facilitates a systematic study of the impacts of multiple training and sampling strategies.  

We perform experimental validation on two popular molecular generation tasks. In protein-small-molecule docking \cite{crampon2022machine}, PharMolixFM-Diffusion achieves competitive performance with AlphaFold3 \cite{abramson2024accurate} (83.4\% v.s. 90.2\%, RMSD < 2Å, given pocket) while enjoying significantly faster inference speed ($\sim$ 4.6 seconds v.s. $\sim$ 249.0 seconds on a single A800 GPU). In structure-based drug design \cite{anderson2003process}, we observe a consistent improvement in the druggability of molecules generated by all PharMolixFM models. Moreover, we explore the empirical inference scaling law of PharMolixFM and perform an in-depth analysis of the impacts of different training tasks.

\section{Related Work}

\subsection{Foundation Models for Unified Molecular Modeling}
Foundation models have been proven successful for numerous data modalities such as texts \cite{touvron2023llama, achiam2023gpt, guo2025deepseek} and images \cite{yuan2021florence, liu2022swin, kirillov2023segment} by achieving outstanding performance on a variety of downstream tasks. However, developing foundation models for biomolecules and their structures remains an open research problem, as different types of biomolecules are composed of different sub-units (e.g. atoms and bonds for small molecules and amino acids for proteins). One line of approach \cite{taylor2022galactica, fang2024mol, luo2024biomedgpt, pei2024biot5+} attempts to transform biomolecular structures into multiple tokens or embedding features and performs large-scale pre-training on a unified language model \cite{vaswani2017attention}. Another line \cite{abramson2024accurate, zheng2024esm, rfaa} proposes to decompose biomolecules into atoms, which are the most fundamental components in structural biology. Then, specialized models and training strategies are designed to capture how these atoms interact with each other to formulate biomolecules in nature. PharMolixFM adopts the all-atom modeling approach to better capture the shared patterns and principles of complex biological systems.

\subsection{Multi-modal Generative Models for Molecules}
Generating structurally plausible molecules is inherently challenging due to the complexity of capturing the joint distribution of both discrete atom types and continuous 3D coordinates. Existing multi-modal generative models address this issue in three ways: (1) Sequence-based models \cite{hayes2025simulating, wang2024dplm} that transform 3D coordinates with vector-quantized tokenizers \cite{van2017neural} and perform generative modeling \cite{sahoo2024simple} on the discrete tokens. (2) Latent generative models \cite{xu2023geometric,fu2024latent,liproteinzen} that encode multi-modal inputs into continuous representations and leverage diffusion \cite{ho2020denoising} or flow matching \cite{lipman2022flow} within the latent space for generative modeling. (3) Fusion models \cite{song2023unified, campbell2024generative} that adopt independent generative processes on each data modality and perform multi-modal fusion at each generation step. PharMolixFM follows the fusion approach and envisions that different downstream tasks can be unified by the same generative denoising process with different priors.

\begin{figure}[htpb]
  \centering 
  \includegraphics[width=0.95\textwidth]{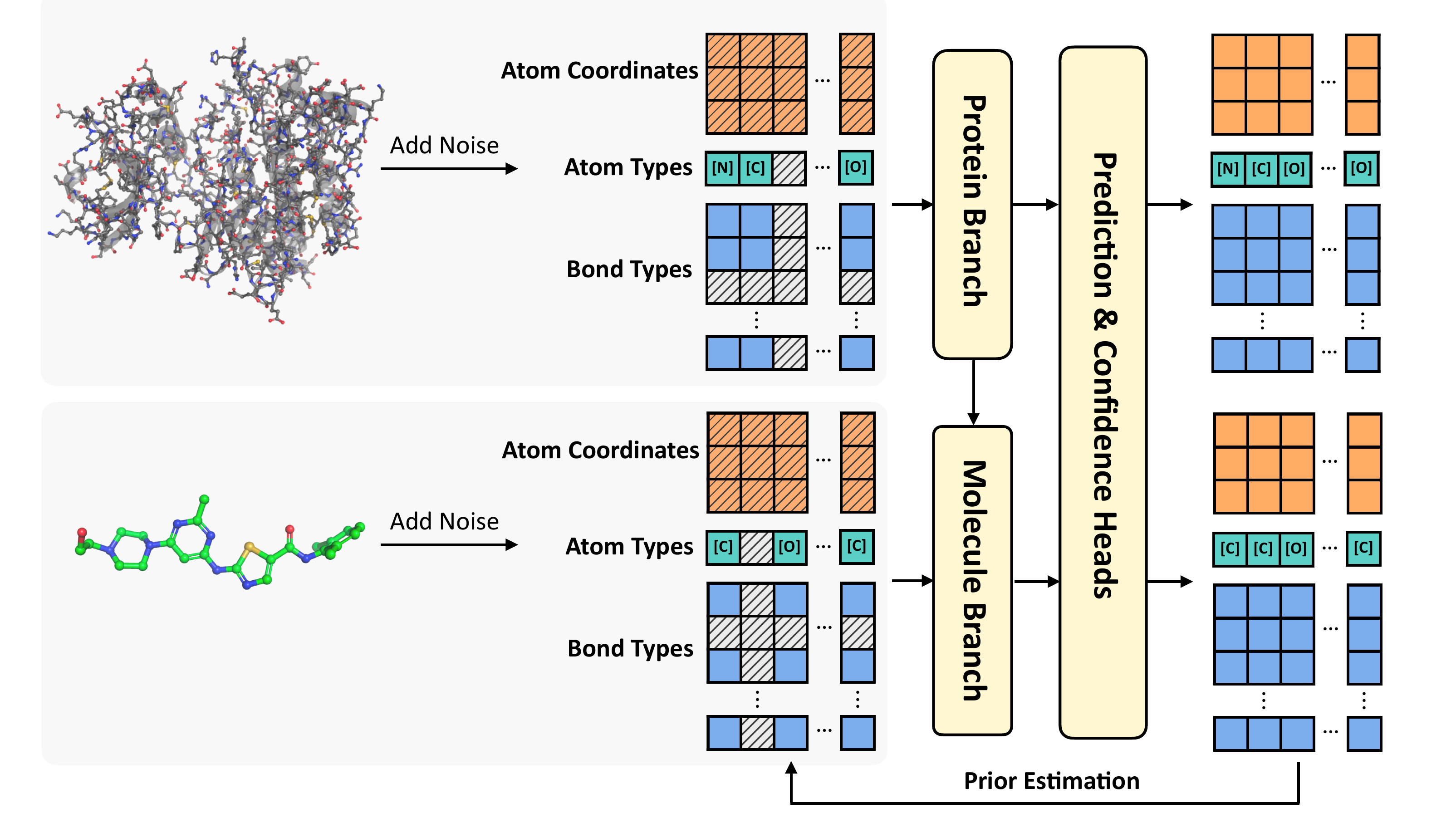} 
  \captionsetup{font={small,stretch=0.95}}
  \caption{\textbf{An overview of the PharMolixFM framework.} It decomposes molecules and proteins into atoms and performs a denoising generative process on atom coordinates, atom types, and bond types. The network architecture composes a protein branch, a molecule branch, and independent prediction heads and confidence heads to reconstruct the original biomolecules.}
  \label{fig:arch} 
\end{figure}

\section{Methods}

\subsection{Preliminaries \label{sec:pre}}
\paragraph{SE-(3) Invariance} In PharMolixFM, we decompose a biomolecular system comprising small molecules, proteins, peptides, or other biological components into $N$ atoms. We formulate its data structure as a triplet $S_0=\langle \mathbf{X}_0, \mathbf{A}_0, \mathbf{E}_0\rangle$ drawn from data distribution $p(S_0)$, where $\mathbf{X}_0\in \mathbb{R}^{N\times 3}$ is the 3-dimensional coordinates for each atom, $\mathbf{A}_0\in \mathbb{Z}^{N\times D_1}$ represents the one-hot encoding of atom type, and $\mathbf{E}_0\in \mathbb{Z}^{N\times N\times D_2}$ is an edge matrix denoting the one-hot bond types between atoms. 

We aim to design a generative model to estimate the underlying density function $p_\theta(S_0)=p_\theta(\langle \mathbf{X}_0, \mathbf{A}_0, \mathbf{E}_0\rangle)$. One of the most important inductive biases of molecules is SE-(3) invariance, \textit{i.e.}, the likelihood should not be influenced by any rotation and translation in 3D geometry. Specifically, given any $3\times 3$ orthogonal matrix $\mathbf{R}$ and translation vector $\mathbf{t}$, the density function should satisfy $p_\theta(\langle \mathbf{R}\mathbf{X}_0+\mathbf{t},\mathbf{A}_0,\mathbf{E}_0\rangle)=p_\theta(\langle\mathbf{X}_0, \mathbf{A}_0, \mathbf{E}_0\rangle)$.

\paragraph{Latent Variable Models}

\begin{figure}[htpb]
  \centering 
  \includegraphics[width=\textwidth]{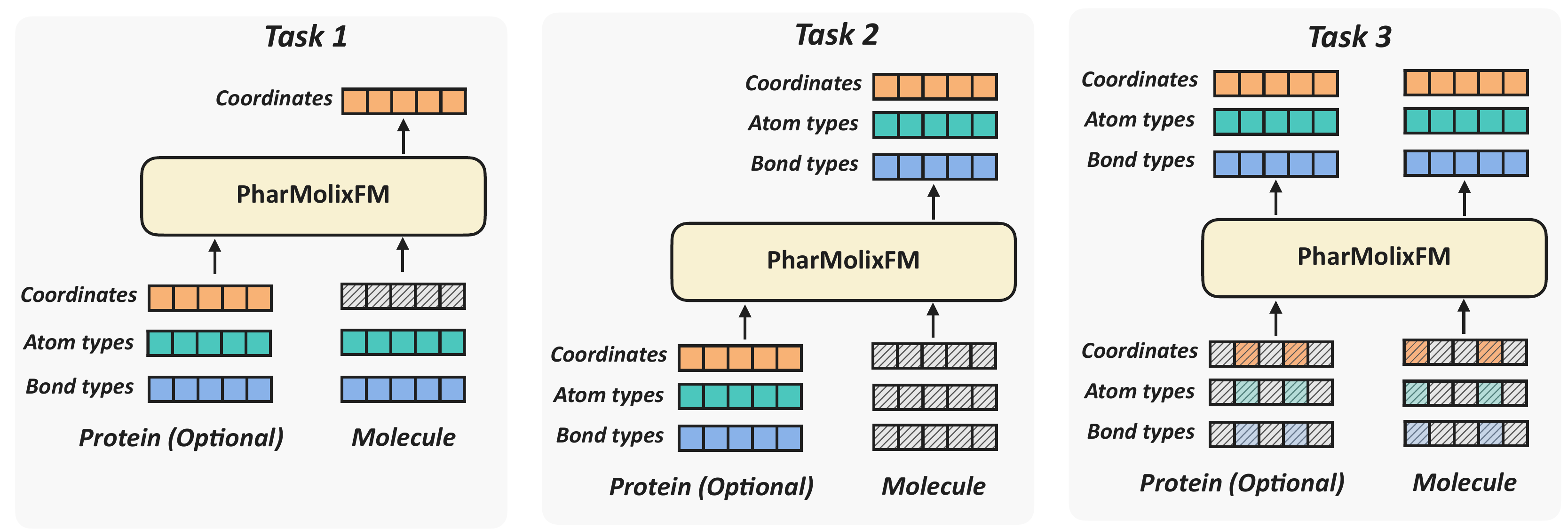} 
  \captionsetup{font={small,stretch=0.95}}
  \caption{\textbf{The training tasks of PharMolixFM.} We introduce three tasks by applying different noises to each variable.} 
  \label{fig:task} 
\end{figure}

We adopt a series of latent variable models to capture the probability distribution $p_\theta(S_0)$. Specifically, we define the latent variables as $S_1,S_2,\cdots,S_T$ which are a series of corrupted versions of $S_0$. Then, we optimize the following variational lower bound:
\begin{equation}
\label{equ:vlb}
\begin{aligned}
    \log p_\theta(S_0)\ge \mathcal{L}_{\text{vlb}}&=\mathbb{E}_{q(S_{1:T}|S_0)}\log\frac{ p_\phi(S_{0:T})}{q(S_{1:T}|S_0)}\\
    &=-D_{KL}(q(S_{1:T}|S_0)\|p_\phi(S_{1:T}))+\mathbb{E}_{q(S_{1:T}|S_0)}\log p_\phi(S_0|S_{1:T})
\end{aligned}
\end{equation}
where $q(S_{1:T}|S_0)$ is variational distribution obtained by gradually adding noise to $S_0$ based on a pre-defined noise schedule $\sigma=(\sigma_1,\sigma_2,\cdots,\sigma_T)$:
\begin{equation}
\label{equ:var_dist}
q(S_{1:T}|S_0)=\prod_{i=1}^T\prod_{j=1}^{N}q\left(\mathbf{X}_{i,j}\big|\mathbf{X}_{0,j},\sigma_{i,j}^{(\mathbf{X})}\right)q\left(\mathbf{A}_{i,j}\big|\mathbf{A}_{0,j},\sigma_{i,j}^{(\mathbf{A})}\right)q\left(\mathbf{E}_{i,j}\big|\mathbf{E}_{0,j},\sigma_{i,j}^{(\mathbf{E})}\right).
\end{equation}
Differing from typical denoising generative models \cite{ho2020denoising, lipman2022flow} where the same noise is applied to each variable, we apply independent noises to different data modalities or atoms to accommodate various downstream tasks. As illustrated in Fig. \ref{fig:task}, in protein-small-molecule docking, we fix the noise as 0 for discrete variables $\textbf{A}$ and $\textbf{E}$ and atom coordinates within the protein, and apply a growing noise to atom coordinates within the molecule. In structure-based drug design, we only apply zero noise to the atoms within the protein. 

Considering the prior $p_\phi$, we assume that it is calculated in an auto-regressive manner, \textit{i.e.} $p_\phi(S_{1:T})=p_\phi(S_T)\prod_{i=1}^{T-1}p_\phi(S_i|S_{i+1:T})$. Following \cite{ho2020denoising}, we parameterize the prior by $p_{\phi}(S_i|S_{i+1:T})\propto \sum_{\tilde{S_0}}q(S_{i:T}|\tilde{S}_0)\tilde{p}_\phi(\tilde{S_0}|S_{i+1:T})$ and use a neural network to predict $\tilde{p}_\phi$ which we detail in Sec. \ref{sec:network}.

During training, we randomly sample $S_0$ from the data distribution and $S_1,S_2,\cdots, S_T$ from the variational distribution in Equation \ref{equ:var_dist}
 and optimize a hybrid objective \cite{nichol2021improved} as follows:

 \begin{equation}
 \label{equ:loss}
 \begin{aligned}
     \mathcal{L}=\mathbb{E}_{q(S_0)}(\mathcal{L}_{\text{vlb}}+\mathbb{E}_{i\sim U(1,T)}\mathbb{E}_{q(S_{i:T}|S_0)}[-\lambda_i\log p_{\phi}(S_0|S_{i:T})])
 \end{aligned}
 \end{equation}
 
During inference, we start with a random sample from the prior $p_\phi(S_T)$ and generate $S_{T-1}, S_{T-2},\cdots, S_1, S_0$ auto-regressively based on the prior $p_\phi$.

\subsection{Network Architecture \label{sec:network}}
To simplify the modeling of the prior, we ensure that the input dimension of the neural network matches the data dimension by either incorporating the Markov assumption into the generation process or performing Bayesian inference on a series of latent parameters. We denote the network inputs as $\tilde{S_i}$ which can be regarded as a noisy version of $S_0$ and the goal of our network is to reconstruct the original data based on noisy inputs. 

We follow the network architecture of PocketXMol \cite{pocketxmol}, which is a dual-branch SE-(3) equivariant graph neural network \cite{satorras2021n} $\Phi$ satisfying the following:

\begin{equation}
    \Phi\left(\langle \tilde{\mathbf{X}}_i,\mathbf{\tilde{A}}_i, \mathbf{\tilde{E}}_i\rangle\right)=\langle \tilde{\mathbf{X}}_0,\mathbf{\tilde{A}}_0, \mathbf{\tilde{E}}_0\rangle,\quad \Phi\left(\langle \mathbf{R}\tilde{\mathbf{X}}_i+\mathbf{t},\mathbf{\tilde{A}}_i, \mathbf{\tilde{E}}_i\rangle\right)=\langle \mathbf{R}\tilde{\mathbf{X}}_0+\mathbf{t},\mathbf{\tilde{A}}_0, \mathbf{\tilde{E}}_0\rangle,
\end{equation}

where $\mathbf{R},\mathbf{t}$ are rotation matrix and translation vector as described in Sec. \ref{sec:pre}.

As illustrated in Fig. \ref{fig:arch}, we first encode atoms within the protein with the protein branch. Then, the representations are passed into the molecule branch to capture the protein-molecule interactions through atoms. Finally, independent prediction heads are used to predict the atom coordinates $\tilde{\mathbf{X}}_0$, atom types $\tilde{\mathbf{A}}_0$, and bond types $\tilde{\mathbf{E}}_0$. We also incorporate prediction heads to estimate the confidence of the prediction. We defer readers to Appendix \ref{app:arch} for a thorough introduction to the network architecture.

\subsection{Multi-modal Generative Models}
PharMolixFM employs 3 types of multi-modal generative models to systematically analyze their effectiveness in modeling biomolecular data. We first define the noise schedule $\sigma_{i,j}^{(\mathcal{M})}=\frac{i}{T}\cdot \text{fix}_{j}^{(\cdot)}$, where $\text{fix}_j^{(\mathcal{M})}$ is a scaling factor ranging from 0 to 1 which control the noise scale of the $j$-th parameter of modality $\mathcal{M}$. We defer readers to Sec. \ref{sec:noise_schedule} for the design of $\text{fix}_j^{(\mathcal{M})}$. Then, we detail the design in noising processes and training objectives of the 3 models as follows.

\paragraph{Multi-modal diffusion} In the diffusion model, the generative process is defined to be Markovian, i.e. $q(S_i|S_0)=\mathbb{E}_{q(S_{i-1}|S_0)}q(S_i|S_{i-1})$ and $\tilde{S_i}=S_{i}$. For atom coordinates, $q(\mathbf{X}_{i,j}|\mathbf{X}_{i-1,j})\triangleq \mathcal{N}(\sqrt{1-\sigma_{i,j}^{(\mathbf{X})}}\mathbf{X}_{i-1,j},\sigma_{i,j}^{(\mathbf{X})}I)$. For atom and bond types, we follow D3PM \cite{austin2021structured} and define the conditional distribution as a categorical distribution by linearly interpolating between a uniform distribution and the one-hot distribution. Through mathematical derivation, we can obtain the following noising process:

\begin{equation}
\label{equ:noise}
\begin{aligned}
&q\left(\mathbf{X}_{i,j}|\mathbf{X}_{0,j},\sigma_{i,j}^{(\mathbf{X})}\right)=\mathcal{N}(\sqrt{\alpha_{i,j}}\ \mathbf{X}_{0,j},(1-\alpha_{i,j})I),\quad \alpha_{i,j}=\prod_{k=1}^{i}\left(1-\sigma_{i,j}^{(\mathbf{X})}\right),\\
&Q_{i,j}^{(\mathbf{A})}=\left((1-\sigma_{i,j}^{(\mathbf{A})})I+\frac{\sigma_{i,j}^{(\mathbf{A})}}{D_1}\mathbbm{1}\mathbbm{1}^T\right),\quad \overline{Q}_{i,j}^{(\mathbf{A})}=\prod_{k=1}^{i}Q_{i,j}^{(\mathbf{A})},\\
&q\left(\mathbf{A}_{i,j}|\mathbf{A}_{0,j},\sigma_{i,j}^{(\mathbf{A})}\right)=\text{Cat}\left(\mathbf{A}_{0,j}\overline{Q}_{i,j}^{(\mathbf{A})}\right),\\
&Q_{i,j}^{(\mathbf{E})}=\left((1-\sigma_{i,j}^{(\mathbf{E})})I+\frac{\sigma_{i,j}^{(\mathbf{E})}}{D_2}\mathbbm{1}\mathbbm{1}^T\right),\quad \overline{Q}_{i,j}^{(\mathbf{E})}=\prod_{k=1}^iQ_{i,j}^{(\mathbf{E})},\\
&q\left(\mathbf{E}_{i,j}|\mathbf{E}_{0,j},\sigma_{i,j}^{(\mathbf{E})}\right)=\text{Cat}\left(\mathbf{E}_{0,j}\overline{Q}_{i,j}^{(\mathbf{E})}\right),
\end{aligned}
\end{equation}

where $\text{Cat}(p)$ denotes the categorical distribution with probabilities given by $p$, $\text{OneHot}(x)$ is the one-hot vector of variable $x$, and $\mathbbm{1}$ is the all-ones column vector. 

The prior is estimated as:
\begin{equation}
\begin{aligned}    &p_\phi(\mathbf{X}_{i,j}|\mathbf{X}_{i+1,j})=\mathcal{N}\left(\frac{\sqrt{1-\sigma_{i+1,j}^{(\mathbf{X})}}(1-\alpha_{i,j})}{1-\alpha_{i+1,j}}\mathbf{X}_{i+1,j}\,+\frac{\sqrt{\alpha_{i,j}}(1-\sigma_{i,j})}{1-\alpha_{i+1,j}}\tilde{\mathbf{X}}_{0,j}, (\sigma_{i,j}^{(\mathbf{X})})^2I\right),\\
&p_\phi(\mathbf{A}_{i,j}|\mathbf{A}_{i+1,j})=\text{Cat}\left(\frac{\mathbf{A}_{i+1,j}(Q_{i+1,j}^{(\mathbf{A})})^T\odot\tilde{\mathbf{A}}_{0,j}\overline{Q}_{i,j}^{(\mathbf{A})}}{\mathbf{A}_{0,j}\overline{Q}_{i+1,j}^{(\mathbf{A})}\mathbf{A}_{i+1,j}^T}\right),\\
&p_\phi(\mathbf{E}_{i,j}|\mathbf{E}_{i+1,j})=\text{Cat}\left(\frac{\mathbf{E}_{i+1,j}(Q_{i+1,j}^{(\mathbf{E})})^T\odot\tilde{\mathbf{E}}_{0,j}\overline{Q}_{i,j}^{(\mathbf{E})}}{\mathbf{E}_{0,j}\overline{Q}_{i+1,j}^{(\mathbf{E})}\mathbf{E}_{i+1,j}^T}\right),
\end{aligned}
\end{equation}

where $\odot$ is the Hadamard product.

The training objective in Equ. \ref{equ:loss} is calculated as follows:
\begin{equation}
    \mathcal{L}=\mathbb{E}_{S_0,i\sim U(1,T),S_i\sim q(S_i|S_0)}\left[\lambda_i^{(\mathbf{X})}\|\tilde{\mathbf{X}}_0-\mathbf{X}_0\|_2^2+\lambda_i^{(\mathbf{A})}\mathcal{L}_{CE}(\tilde{\mathbf{A}}_0,\mathbf{A}_0)+\lambda_i^{(\mathbf{E})}\mathcal{L}_{CE}(\tilde{\mathbf{E}}_0,\mathbf{E}_0)\right],
\end{equation}
where $\|\cdot\|_2$ denotes the $l_2$ norm, and $\mathcal{L}_{CE}$ denotes cross entropy.

\paragraph{Multi-modal flow matching} Flow matching aims to construct a direct mapping from the data distribution \( p(S_0) \) to the prior \( p(S_T) \). The key idea is to learn a time-dependent velocity field $u(S_i)$ that determines a probability path between these distributions based on the continuity equation. Due to the intractability of the marginal vector field, the flow matching model typically optimizes an equivalent form, \textit{i.e.} the conditional vector field $u(S_i|S_0)$. Following \cite{campbell2024generative}, we define the conditional probability of the atom coordinates as the Gaussian path and the conditional probability of the atom and bond types as a Markov chain. Specifically:

\begin{equation}
\begin{aligned}   
&q(\mathbf{X}_{i,j} \mid \mathbf{X}_{0,j}, \sigma_{i,j}^{(\mathbf{X})}) = \mathcal{N} \left( (1-\sigma_{i,j}^{(\mathbf{X})})\mathbf{X}_{0,j}, (\sigma_{i,j}^{(X)})^2I \right), \\
&q(\mathbf{A}_{i,j} \mid \mathbf{A}_{0,j}, \sigma_{i,j}^{(\mathbf{A})}) = \text{Cat}(\mathbf{A}_{0,j}Q_{i,j}^{(\mathbf{A})}),  \\
&q(\mathbf{E}_{i,j} \mid \mathbf{E}_{0,j}, \sigma_{i,j}^{(\mathbf{E})}) =\text{Cat}(\mathbf{E}_{0,j}Q_{i,j}^{(\mathbf{E})}),
\end{aligned}
\end{equation}

where $Q_{i,j}^{(\mathbf{A})}$ and $Q_{i,j}^{(\mathbf{E})}$ are derived in Equ. \ref{equ:noise}. It's worth noting here that these two matrices are used to define the conditional flow instead of the forward diffusion process. 

The conditional vector fields for each modality are defined as follows:
\begin{equation}
\begin{aligned}
    &u(\mathbf{X}_{i,j}|\mathbf{X}_{0,j})=\frac{\mathbf{X}_{0,j}-\mathbf{X}_{i,j}}{\sigma_{i,j}^{(\mathbf{X})}},\\
    &u(\mathbf{A}_{i,j}|\mathbf{A}_{0,j})=\mathbf{A}_{i,j}\frac{\mathbf{A}_{0,j}(\mathbbm{1}-\mathbf{A}_{0,j}^T)}{\sigma_{i,j}^{(\mathbf{A})}},\\
    &u(\mathbf{E}_{i,j}|\mathbf{E}_{0,j})=\mathbf{E}_{i,j}\frac{\mathbf{E}_{0,j}(\mathbbm{1}-\mathbf{E}_{0,j}^T)}{\sigma_{i,j}^{(\mathbf{E})}}.
\end{aligned}
\end{equation}

During training, the neural network $u_\phi(S_i|\sigma_i)$ aims to predict the conditional vector field based on $S_i$. Then, we can perform sampling by solving the following ordinary differential equation (ODE) by the Euler method: 

\begin{equation}
\begin{aligned}
    &p_\phi(\mathbf{X}_{i,j}|\mathbf{X}_{i+1,j})=\mathbf{X}_{i+1,j}+\frac{1}{T}u_\phi(\mathbf{X}_{i+1,j}),\\
    &p_\phi(\mathbf{A}_{i,j}|\mathbf{A}_{i+1,j})=\text{Cat}\left(\mathbf{A}_{i+1,j}+\frac{1}{T}u_\phi(\mathbf{A}_{i+1,j})\right),\\
    &p_\phi(\mathbf{E}_{i,j}|\mathbf{E}_{i+1,j})=\text{Cat}\left(\mathbf{E}_{i+1,j}+\frac{1}{T}u_\phi(\mathbf{E}_{i+1,j})\right).\\
\end{aligned}
\end{equation}
The training objective in Equ. \ref{equ:loss} is calculated as follows:
\begin{equation}
    \begin{aligned}
        \mathcal{L} = \mathbb{E}_{S_0,i\sim U(1,T),S_i\sim q(S_i|S_0)}
        \Bigg[
        &\lambda_i^{(\mathbf{X})} \left\| u_\phi(\mathbf{X}_{i,j}) - (\mathbf{X}_{0,j} - \mathbf{X}_{T,j}) \right\|_2^2 \\
        &+ \lambda_i^{(\mathbf{A})} \mathcal{L}_{CE}(u_\phi(\mathbf{A}_{i,j}),\mathbf{A}_0)
        + \lambda_i^{(\mathbf{E})} \mathcal{L}_{CE}(u_\phi(\mathbf{E}_{i,j}),\mathbf{E}_0)
        \Bigg].
    \end{aligned}
\end{equation}

\paragraph{Bayesian flow networks} The key idea of Bayesian flow networks (BFNs) is to perform generative modeling by a sender-receiver framework within the parameter space of the data distribution rather than in the data space. In BFNs, the posterior $q$ is defined to ensure that the information increases linearly along the generative process. The noisy samples are transformed into the parameter space, \textit{i.e.} the mean and variance of the Gaussian distributions for atom coordinates and the probability of categorical distributions for atom types and bond types. Then, a Bayesian update step is performed on the parameters, which are further fed into the neural network to model their interdependencies. Finally, the parameters are transformed back into the sample space to reconstruct the original data and its noisy version.

The distribution of the Bayesian flow network inputs, known as the \textit{Bayesian flow distribution}, is derived as follows:
\begin{equation}
\begin{aligned}
    &\gamma_{i,j}=1-\alpha^{2(1-\sigma_{i,j}^{(\mathbf{X})})},\ \beta_{i,j}^{(\mathbf{A})}=\beta^{(1-\sigma_{i,j}^{(\mathbf{A})})^2},\ \beta_{i,j}^{(\mathbf{E})}=\beta^{(1-\sigma_{i,j}^{(\mathbf{E})})^2}\\
    & p_F\left(\tilde{\mathbf{X}}_{i,j}^{(\theta)}|\mathbf{X}_{0,j}\right)= \mathcal{N}\left(\gamma_{i,j}\mathbf{X}_{0,j},\gamma_{i,j}(1-\gamma_{i,j})I\right),\\
    & p_F\left(\tilde{\mathbf{A}}_{i,j}^{(\theta)}|\mathbf{A}_{0,j}\right)= \mathcal{N}\left(\beta_{i,j}^{(\mathbf{A})}(D_1\mathbf{A}_{0,j}-1),\beta_{i,j}^{(\mathbf{A})}D_1I\right),\\
    & p_F\left(\tilde{\mathbf{E}}_{i,j}^{(\theta)}|\mathbf{E}_{0,j}\right)= \mathcal{N}\left(\beta_{i,j}^{(\mathbf{E})}(D_2\mathbf{E}_{0,j}-1),\beta_{i,j}^{(\mathbf{E})}D_2I\right),
\end{aligned}
\end{equation}
where $\alpha$ and $\beta$ are pre-defined hyper-parameters. 

Given network outputs $\tilde{\mathbf{X}}_0^{(\theta)},\tilde{\mathbf{A}}_0^{(\theta)},\tilde{\mathbf{E}}_0^{(\theta)}$, the BFN model reconstructs the original data following the \textit{output distribution}:

\begin{equation}
\begin{aligned}
    &p_O\left(\tilde{\mathbf{X}}_{0,j}|\tilde{\mathbf{X}}_{i,j}^{(\theta)}\right)=\mathcal{N}\left(\tilde{\mathbf{X}}_{i,j}^{(\theta)},-\frac{\alpha^{2(1-\sigma_{i,j}^{(\mathbf{X})})}}{2\ln\alpha}I\right),\\
    &p_O(\tilde{\mathbf{A}}_{0,j}|\tilde{\mathbf{A}}_{i,j}^{(\theta)})=\text{Cat}(\text{SoftMax}(\tilde{\mathbf{A}}_{i,j}^{(\theta)})),\\
    &p_O(\tilde{\mathbf{E}}_{0,j}|\tilde{\mathbf{E}}_{i,j}^{(\theta)})=\text{Cat}(\text{SoftMax}(\tilde{\mathbf{E}}_{i,j}^{(\theta)})),
\end{aligned}
\end{equation}

where $\text{SoftMax}$ is the softmax function.

Following \cite{molcraft}, we skip the sampling step and perform Bayesian updates directly within the parameter space to reduce noise during the generation process. Specifically, , the \textit{update distribution} is calculated as follows:
\begin{equation}
\begin{aligned}
   &p_U\left(\tilde{\mathbf{X}}_{i,j}^{(\theta)}\big| \tilde{\mathbf{X}}_{i+1,j}^{(\theta)}\right)=\mathcal{N}\left(\gamma_{i,j}\tilde{\mathbf{X}}_{i+1,j}^{(\theta)},\gamma_{i,j}(1-\gamma_{i,j})I\right),\\
& p_U\left(\tilde{\mathbf{A}}_{i,j}^{(\theta)}\big|\tilde{\mathbf{A}}_{i+1,j}^{(\theta)}\right)= \mathbb{E}_{y\sim \mathcal{N}(\beta_{i,j}^{(\mathbf{A})}(D_1\tilde{\mathbf{A}}_{i,j}-1),\beta_iD_1I)}\delta\left(\tilde{\mathbf{A}}_{i,j}^{(\theta)}-\text{SoftMax(y)}\right),\\
&p_U\left(\tilde{\mathbf{E}}_{i,j}^{(\theta)}\big|\tilde{\mathbf{E}}_{i+1,j}^{(\theta)}\right)= \mathbb{E}_{y\sim \mathcal{N}(\beta_{i,j}^{(\mathbf{E})}(D_2\tilde{\mathbf{E}}_{i,j}-1),\beta_iD_2I)}\delta\left(\tilde{\mathbf{E}}_{i,j}^{(\theta)}-\text{SoftMax(y)}\right).
\end{aligned}
\end{equation}

The discrete-time training objective of the BFN model in Equ. \ref{equ:loss} is derived in the following:
\begin{equation}
\begin{aligned}
    &\mathcal{L}=\mathbb{E}_{S_0,i\sim U(1,T),\tilde{S}_i\sim p_F(\tilde{S}_i^{(\theta)}|S_0)}\left[\lambda_i^{(\mathbf{X})}\|\mathbf{X}_i-\tilde{\mathbf{X}}_{i}^{(\theta)}\|_2^2+\right.\\ &\left.\qquad \qquad \qquad \qquad \qquad \qquad \  \mathcal{L}_{d}(\mathbf{A}_i^{(\theta)},\mathbf{\tilde{A}}_{i}^{(\theta)},D_1,\beta_{i}^{(\mathbf{A})})+\mathcal{L}_{d}(\mathbf{E}_i^{(\theta)},\mathbf{\tilde{E}}_{i}^{(\theta)},D_2,\beta_{i}^{(\mathbf{E})})\right],\\
    &\mathcal{L}_d(\mathbf{A},\tilde{\mathbf{A}}, D,\beta)=\sum_{j=1}^{N}[\ln \mathcal{N}\left(y|\beta_j(K\mathbf{A}_j-1),\beta_jKI\right)\\ & \qquad \qquad \qquad \qquad \ -\ln \sum_{k=1}^{D}\text{SoftMax}(\tilde{\mathbf{A}}_j)\mathcal{N}(y|\beta_j(K\mathbf{e}_j-1),\beta_jKI)].
\end{aligned}
\end{equation}

\subsection{Enhancing Multi-Task Transferability with Mixed Denoising Processes \label{sec:noise_schedule}}

Inspired by PocketXMol \cite{pocketxmol}, we introduce a $\text{Fix}$ term to control the noise scale and prior distribution for each atom and modality. When $\text{Fix}=1$, the noising process is the same as the original generative model. When $\text{Fix}=0$, the corresponding variable remains unchanged and the denoising process could be viewed as a conditional generation process. This facilitates the joint training of multiple tasks within a unified framework. Besides, $\text{Fix}$ can take an arbitrary value between $0$ and $1$ to represent the confidence of the prior, \textit{e.g.}, the molecule atoms are located close to a pocket with a certain size. As shown in Fig. \ref{fig:task}, we incorporate multiple denoising processes for training to enhance the transferability of our model by introducing different $\text{Fix}$ values, detailed as follows:
\begin{itemize}[noitemsep,topsep=0pt,parsep=5pt,partopsep=0pt,leftmargin=20pt]
    \item $\text{Fix}=1$ for the protein atoms and atom and bond types of small-molecule atoms, and $\text{Fix}=0$ for the coordinates of small-molecule atoms. This task predicts the molecular conformation based on a binding protein (if available) and its 2D molecular graph, which can be applied in protein-small-molecule docking \cite{crampon2022machine} and molecular conformation prediction \cite{xugeodiff}. 
    \item $\text{Fix}=1$ for the protein atoms and $\text{Fix}=0$ for the molecule atoms. This task aims to generate a novel molecule based on a binding protein (if available), which is common in denovo molecule generation \cite{gebauer2019symmetry} and structure-based drug design \cite{austin2021structured}. 
    \item $\text{Fix}=0.7$ for 15\% of randomly selected atoms and $\text{Fix}=0$ for the remaining atoms. This task simulates the process of first determining the most crucial component within the biomolecular system and then predicting the other components. This enhances the robustness of our model and facilitates downstream applications such as fragment linking \cite{ichihara2011compound} and antibody design \cite{kong2022conditional}.  
\end{itemize}

\begin{table}[htpb]
\centering
\captionsetup{font={small,stretch=0.95}}
\caption{\textbf{Performance on protein-small-molecule docking task.} The RMSD values for DiffDock and Vina are taken from PoseBusters \cite{posebusters}, and the RMSD values for the other baselines are derived from the original paper. The best results are marked bold, and the second-best results are underlined.}
\begin{tabular}{lcc}
\toprule
\multirow{2}{*}{\textbf{Method}} & \multicolumn{2}{c}{\textbf{Ratio of RMSD < 2.0Å (\%)}} \\
 \cmidrule(lr){2-3} 
 & \textbf{self-ranking} & \textbf{oracle-ranking} \\
\midrule
DiffDock \cite{diffdock} & 38.0 & - \\
RFAA \cite{rfaa} & 42.0 & - \\
Vina \cite{posebusters} & 52.3 & - \\
UniMol-Docking V2 \cite{unimoldockingv2} & 77.6 & - \\
SurfDock \cite{surfdock} & 78.0 & - \\
AlphaFold3 \cite{abramson2024accurate} & \textbf{90.4} & - \\
PocketXMol (50 repeats) \cite{pocketxmol} & 82.2 & 95.3 \\ \midrule
PharMolixFM-Diff (50 repeats) & 83.4 & 96.0 \\
PharMolixFM-Flow (50 repeats) & 73.4 & 93.7\\
PharMolixFM-BFN (50 repeats) & 78.5 & 93.5 \\ \midrule
PharMolixFM-Diff (500 repeats) & \underline{83.9} & \textbf{98.1} \\
PharMolixFM-Flow (500 repeats) & 74.8 & 96.7 \\
PharMolixFM-BFN (500 repeats) & 78.5 & \underline{97.9} \\
\bottomrule
\end{tabular}
\label{tab:docking}
\end{table}

\section{Experiments}

In this section, we demonstrate that PharMolixFM is adept at molecular modeling and generation. We start with a brief introduction of our training setups, followed by detailed evaluations on the two popular molecular generation tasks: protein-small-molecule docking and structure-based drug design. We also explore the inference scaling law and the impacts of different training tasks.

\subsection{Training Setup}
We train PharMolixFM with a batch size of 40 for 180K steps on four 80G A800 GPUs. For each sample, the training task is selected with equal probability. We adopt the AdamW \cite{loshchilov2019decoupled} optimizer with a weight decay of $\lambda=0.001$, $\beta_{1}=0.99$ and $\beta_{2}=0.999$. The learning rate is linearly warmed up to $0.001$ for the first 1000 steps. 

\paragraph{Dataset.}
The training dataset is collected from multiple databases or datasets, including PDBBind \cite{pdbbind}, Binding MOAD \cite{moad}, CrossDocked2020 \cite{crossdocked}, and PepBDB dataset \cite{pepbdb}. We adopt the same data filtering strategy as PocketXMol \cite{pocketxmol}. The molecule should be successfully loaded by RDKit and include only pre-defined atom types, like C, N, O, F, P, S, Cl, B, Br, I, and Se. The count of heavy atoms should be within the range of 4 to 122. For peptides, the length should be smaller than 15 and the count of heavy atoms should not exceed 150.

\subsection{Performance Evaluation on Protein-small-molecule Docking}
Protein-small-molecule docking aims to predict the binding pose of a receptor-ligand complex. We adopt the holo docking \cite{diffdock} setting where the ligand-bound structure of the protein is known and fixed. We also provide the model with a 10Å binding pocket based on the reference ligand.

\paragraph{Evaluation.}
We use the PoseBusters benchmark \cite{posebusters}, which consists of 428 protein-small-molecule complex structures. Following the benchmark setting, we measure the RMSD (Root-Mean-Square Deviation) value between the predicted binding pose and the ground-truth binding pose and report the ratio of RMSD < 2Å among the 428 complex structures. Following PocketXMol, we perform docking multiple times for each test sample and select one docking pose based on self-ranking or oracle-ranking. For self-ranking, the generated poses are ranked by the average confidence scores at each step of the generation process. For oracle-ranking, we use the RMSD scores with the ground-truth binding pose as the ranking score. It's worth noting that in practice the ground-truth binding pose is unavailable, and the oracle-ranking setting is only used to investigate if the model is able to produce a satisfying result based on repeated sampling.

\paragraph{Baselines.}
We adopt DiffDock \cite{diffdock}, Vina \cite{posebusters}, RFAA \cite{rfaa}, UniMol-Docking V2 \cite{unimoldockingv2}, SurfDock \cite{surfdock}, PocketXMol \cite{pocketxmol}, and AlphaFold3 \cite{abramson2024accurate} as baselines. 

\paragraph{Results and analysis.}

 The results of protein-small-molecule docking are displayed in Tab.~\ref{tab:docking}. The key observations include: (1) ParMolixFM-Diff achieves the second-best results under the self-ranking setting, outperforming PocketXMol by 1.7\% absolute gains. While PharMolixFM-Diff underperforms AlphaFold3, our model enjoys significantly faster inference speed ($\sim$ 4.6 seconds v.s. $\sim$ 249.0 seconds on a single A800 GPU). (2) Under the oracle-ranking setting, the performance can be boosted to 98.0\% if the best docking pose is selected. This corroborates prior studies \cite{pocketxmol, corso2024deep} and indicates that advanced design in ranking scores and sampling strategies may further improve the docking performance. (3) PharMolixFM-BFN slightly underperforms PharMolixFM-Diff, and we attribute this to their difference in noising processes. At early sampling steps, the noise scales for BFN are much smaller than those in the diffusion model, making the training task less challenging. (4) PharMolixFM-Flow takes more sampling repeats to achieve competitive performance with the BFN and diffusion counterparts. Since the sampling process of flow matching models is deterministic, we speculate that introducing certain noise terms during the generation process may be beneficial.

\subsection{Performance Evaluation on Structure-based Drug Design}
Differing from the docking task where the atom and bond types of the small molecule is given, structure based drug design (SBDD) aims to directly generate a small molecule based on a given pocket.

\paragraph{Evaluation.}

We use the test set of CrossDocked \cite{crossdocked}, which is a widely-adopted dataset comprising 100 protein pocket. For each pocket, we generate 100 molecules and report their binding affinity scores and drug-like properties. Specifically, we employ AutoDock Vina \cite{trott2010autodock} to measure binding affinity and report Vina score (score of the designed molecule), Vina min (score of the optimized pose after a local minimization of energy), and Vina dock (the best possible score after re-docking). We use QED (Quantitative Estimation of Drug-likeness) and SA (Synthetic Accessibility) to evaluate the druggability of the molecule. 

\paragraph{Baselines.}
We conduct a comprehensive comparison against representative methods, including an autoregressive sampling-based method Pocket2Mol \cite{pocket2mol}, two diffusion-based methods TargetDiff \cite{targetdiff} and DecompDiff \cite{decompdiff}, and a BFN-based method MolCRAFT \cite{molcraft}.


\begin{table}[htpb]
    \centering
    \captionsetup{font={small,stretch=0.95}}
    \caption{\textbf{Performance on structure-based drug design.} The baseline results are derived from the original paper. The best results are marked bold, and the second-best results are underlined.}
    \renewcommand{\arraystretch}{1.2}
    \begin{tabular}{cccccc}
        \toprule
        \multirow{3}{*}{\textbf{Method}} & \multicolumn{3}{c}{\textbf{Binding Affinity}} & \multicolumn{2}{c}{\textbf{Drug-like Property}} \\
        \cmidrule(lr){2-4} \cmidrule(lr){5-6}
        & \textbf{Vina score↓} & \textbf{Vina min↓} & \textbf{Vina dock↓} & \textbf{QED↑} & \textbf{SA↑} \\
        & (Avg./Med.) & (Avg./Med.) & (Avg./Med.) & (Avg.) & (Avg.) \\
        \midrule
        Pocket2Mol \cite{pocket2mol}   & -5.14/-4.70 & -6.42/-5.82 & -7.15/6.79  & \textbf{0.57}  & \textbf{0.76} \\
        TargetDiff \cite{targetdiff}   & -5.47/-6.30 & -6.64/-6.83 & -7.80/-7.91 & 0.48  & 0.58 \\
        DecompDiff \cite{decompdiff}   & -5.67/-6.04 & -7.04/-7.09 & \underline{-8.39/-8.43} & 0.45  & 0.61 \\
        MolCRAFT \cite{molcraft}   & \textbf{-6.61/-8.14} & \textbf{-8.14/-8.42} & \textbf{-9.25/-9.20} & 0.46  & 0.62 \\ \midrule
        PharMolixFM-Diff & -6.18/-6.44 & \underline{-7.19/-7.04} & -7.88/-7.85 & \underline{0.50}  & 0.73 \\
        PharMolixFM-Flow & -6.34/\underline{-6.47} & -7.17/-7.06 & -7.87/-7.82 & 0.49 & \underline{0.74} \\
        PharMolixFM-BFN  & \underline{-6.38}/-6.45 & -6.97/-6.89 & -7.68/-7.64 & 0.48  & 0.64 \\
        \bottomrule
    \end{tabular}
    \label{tab:sbdd}
\end{table}

\paragraph{Results and analysis.}

We perform comparisons on CrossDocked dataset in Tab.~\ref{tab:sbdd}. We observe that: (1) PharMolixFM achieves a better balance between binding affinity and drug-like properties. PharMolixFM-Diff outperforms the diffusion-based methods TargetDiff and DecompDiff on vina score, QED, and SA. Besides, PharMolixFM-BFN achieves -6.38 average vina score, only second to the state-of-the-art method MolCRAFT. While improved vina scores are generally favored in this task, it is worth noting that the druggability of the molecule is also significant for further validation and application in vivo \cite{gao2024theory}. Therefore, we argue that PharMolixFM can serve as an alternative in practical drug design scenarios. (2) The performance of different PharMolixFM variants exhibit little difference, indicating that the training data and model architecture may play a more important role in this task.


\subsection{PharMolixFM Exhibits the Empirical Inference Scaling Law\label{sec:inf_scaling}}

\begin{figure}
  \centering 
  \includegraphics[width=\textwidth]{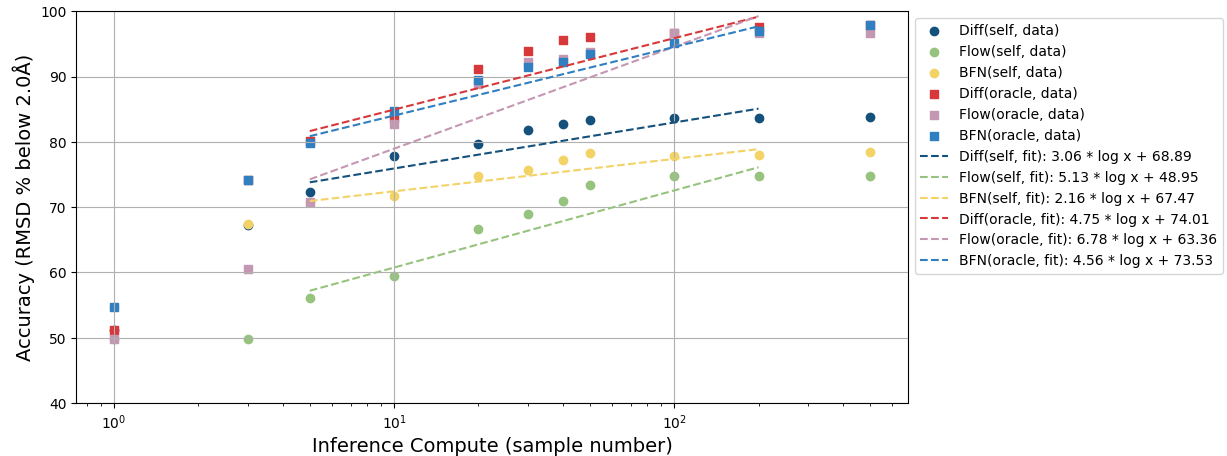} 
    \captionsetup{font={small,stretch=0.95}}
  \caption{\textbf{Investigations on the inference scaling law of PharMolixFM.} We show the accuracy on the protein-small-molecule docking task for all PharMolixFM models and the fitted scaling curve using least square method.} 
  \label{fig:your_label} 
\end{figure}

Inspired by the recent success of inference scaling in the NLP domain \cite{wu2025inference, shao2024deepseekmath, yu2025dapo}, we investigate if the performance of molecular generative models could be improved by increasing compute during inference time. We perform analysis on the protein-small-molecule docking task and assume that the ratio of RMSD < 2Å has the following correlations with the sampling repeats $R$, which is proportional to inference time:
\begin{equation}
\text{Acc}=a \log (b R+c)+d,    
\end{equation}
where $a,b,c,d$ are constants.

As shown in Figure 1, we observe that within a certain range, the performance could be improved by a certain degree if the inference compute is doubled, indicating that the empirical inference scaling law holds for all PharMolixFM models. It is important to note that due to the distributional differences between the training and test sets, the model performance will not increase indefinitely with additional sampling. This shed light on the potential power of foundation models in molecular structures and future possibilities of the \textit{Aha! moment} for structural biology. 


\section{Conclusions}
In this work, we present PharMolixFM, a unified framework for developing all-atom foundation models for molecular modeling and generation. We implement 3 variants of PharMolixFM using state-of-the-art multi-modal generative models and propose novel training tasks to enhance the multi-task transferability of our models. Experiments show that PharMolixFM models achieve state-of-the-art performance on protein-small-molecule docking and structure-based drug-design while enjoying fast inference speed. More importantly, we explore the inference scaling law of PharMolixFM models, which is one of the most significant properties of foundation models. With responsible and ethical use, PharMolixFM bears promise in accelerating research in life sciences and drug discovery.

\bibliographystyle{unsrt}
\bibliography{ref}

\begin{thebibliography}{10}

\bibitem{congreve2005keynote}
Miles Congreve, Christopher~W Murray, and Tom~L Blundell.
\newblock Keynote review: Structural biology and drug discovery.
\newblock {\em Drug discovery today}, 10(13):895--907, 2005.

\bibitem{whisstock2003prediction}
James~C Whisstock and Arthur~M Lesk.
\newblock Prediction of protein function from protein sequence and structure.
\newblock {\em Quarterly reviews of biophysics}, 36(3):307--340, 2003.

\bibitem{yan2005mechanisms}
Nieng Yan and Yigong Shi.
\newblock Mechanisms of apoptosis through structural biology.
\newblock {\em Annu. Rev. Cell Dev. Biol.}, 21(1):35--56, 2005.

\bibitem{surade2012structural}
Sachin Surade and Tom~L Blundell.
\newblock Structural biology and drug discovery of difficult targets: the limits of ligandability.
\newblock {\em Chemistry \& biology}, 19(1):42--50, 2012.

\bibitem{jumper2021highly}
John Jumper, Richard Evans, Alexander Pritzel, Tim Green, Michael Figurnov, Olaf Ronneberger, Kathryn Tunyasuvunakool, Russ Bates, Augustin {\v{Z}}{\'\i}dek, Anna Potapenko, et~al.
\newblock Highly accurate protein structure prediction with alphafold.
\newblock {\em nature}, 596(7873):583--589, 2021.

\bibitem{baek2021accurate}
Minkyung Baek, Frank DiMaio, Ivan Anishchenko, Justas Dauparas, Sergey Ovchinnikov, Gyu~Rie Lee, Jue Wang, Qian Cong, Lisa~N Kinch, R~Dustin Schaeffer, et~al.
\newblock Accurate prediction of protein structures and interactions using a three-track neural network.
\newblock {\em Science}, 373(6557):871--876, 2021.

\bibitem{dauparas2022robust}
Justas Dauparas, Ivan Anishchenko, Nathaniel Bennett, Hua Bai, Robert~J Ragotte, Lukas~F Milles, Basile~IM Wicky, Alexis Courbet, Rob~J de~Haas, Neville Bethel, et~al.
\newblock Robust deep learning--based protein sequence design using proteinmpnn.
\newblock {\em Science}, 378(6615):49--56, 2022.

\bibitem{burley2017protein}
Stephen~K Burley, Helen~M Berman, Gerard~J Kleywegt, John~L Markley, Haruki Nakamura, and Sameer Velankar.
\newblock Protein data bank (pdb): the single global macromolecular structure archive.
\newblock {\em Protein crystallography: methods and protocols}, pages 627--641, 2017.

\bibitem{varadi2022alphafold}
Mihaly Varadi, Stephen Anyango, Mandar Deshpande, Sreenath Nair, Cindy Natassia, Galabina Yordanova, David Yuan, Oana Stroe, Gemma Wood, Agata Laydon, et~al.
\newblock Alphafold protein structure database: massively expanding the structural coverage of protein-sequence space with high-accuracy models.
\newblock {\em Nucleic acids research}, 50(D1):D439--D444, 2022.

\bibitem{lin2023evolutionary}
Zeming Lin, Halil Akin, Roshan Rao, Brian Hie, Zhongkai Zhu, Wenting Lu, Nikita Smetanin, Robert Verkuil, Ori Kabeli, Yaniv Shmueli, et~al.
\newblock Evolutionary-scale prediction of atomic-level protein structure with a language model.
\newblock {\em Science}, 379(6637):1123--1130, 2023.

\bibitem{awais2025foundation}
Muhammad Awais, Muzammal Naseer, Salman Khan, Rao~Muhammad Anwer, Hisham Cholakkal, Mubarak Shah, Ming-Hsuan Yang, and Fahad~Shahbaz Khan.
\newblock Foundation models defining a new era in vision: a survey and outlook.
\newblock {\em IEEE Transactions on Pattern Analysis and Machine Intelligence}, 2025.

\bibitem{yuan2021florence}
Lu~Yuan, Dongdong Chen, Yi-Ling Chen, Noel Codella, Xiyang Dai, Jianfeng Gao, Houdong Hu, Xuedong Huang, Boxin Li, Chunyuan Li, et~al.
\newblock Florence: A new foundation model for computer vision.
\newblock {\em arXiv preprint arXiv:2111.11432}, 2021.

\bibitem{zhou2024comprehensive}
Ce~Zhou, Qian Li, Chen Li, Jun Yu, Yixin Liu, Guangjing Wang, Kai Zhang, Cheng Ji, Qiben Yan, Lifang He, et~al.
\newblock A comprehensive survey on pretrained foundation models: A history from bert to chatgpt.
\newblock {\em International Journal of Machine Learning and Cybernetics}, pages 1--65, 2024.

\bibitem{touvron2023llama}
Hugo Touvron, Thibaut Lavril, Gautier Izacard, Xavier Martinet, Marie-Anne Lachaux, Timoth{\'e}e Lacroix, Baptiste Rozi{\`e}re, Naman Goyal, Eric Hambro, Faisal Azhar, et~al.
\newblock Llama: Open and efficient foundation language models.
\newblock {\em arXiv preprint arXiv:2302.13971}, 2023.

\bibitem{abramson2024accurate}
Josh Abramson, Jonas Adler, Jack Dunger, Richard Evans, Tim Green, Alexander Pritzel, Olaf Ronneberger, Lindsay Willmore, Andrew~J Ballard, Joshua Bambrick, et~al.
\newblock Accurate structure prediction of biomolecular interactions with alphafold 3.
\newblock {\em Nature}, 630(8016):493--500, 2024.

\bibitem{ho2020denoising}
Jonathan Ho, Ajay Jain, and Pieter Abbeel.
\newblock Denoising diffusion probabilistic models.
\newblock {\em Advances in neural information processing systems}, 33:6840--6851, 2020.

\bibitem{rfaa}
Rohith Krishna, Jue Wang, Woody Ahern, Pascal Sturmfels, Preetham Venkatesh, Indrek Kalvet, Gyu~Rie Lee, Felix~S Morey-Burrows, Ivan Anishchenko, Ian~R Humphreys, et~al.
\newblock Generalized biomolecular modeling and design with rosettafold all-atom.
\newblock {\em Science}, 384(6693):eadl2528, 2024.

\bibitem{zheng2024esm}
Kangjie Zheng, Siyu Long, Tianyu Lu, Junwei Yang, Xinyu Dai, Ming Zhang, Zaiqing Nie, Wei-Ying Ma, and Hao Zhou.
\newblock Esm all-atom: multi-scale protein language model for unified molecular modeling.
\newblock In {\em Proceedings of the 41st International Conference on Machine Learning}, pages 61432--61453, 2024.

\bibitem{austin2021structured}
Jacob Austin, Daniel~D Johnson, Jonathan Ho, Daniel Tarlow, and Rianne Van Den~Berg.
\newblock Structured denoising diffusion models in discrete state-spaces.
\newblock {\em Advances in neural information processing systems}, 34:17981--17993, 2021.

\bibitem{campbell2024generative}
Andrew Campbell, Jason Yim, Regina Barzilay, Tom Rainforth, and Tommi Jaakkola.
\newblock Generative flows on discrete state-spaces: enabling multimodal flows with applications to protein co-design.
\newblock In {\em Proceedings of the 41st International Conference on Machine Learning}, pages 5453--5512, 2024.

\bibitem{graves2023bayesian}
Alex Graves, Rupesh~Kumar Srivastava, Timothy Atkinson, and Faustino Gomez.
\newblock Bayesian flow networks.
\newblock {\em arXiv preprint arXiv:2308.07037}, 2023.

\bibitem{pocketxmol}
Xingang Peng, Ruihan Guo, Yan Xu, Jiaqi Guan, Yinjun Jia, Yanwen Huang, Muhan Zhang, Jian Peng, Jiayu Sun, Chuanhui Han, et~al.
\newblock Decipher fundamental atomic interactions to unify generative molecular docking and design.
\newblock {\em bioRxiv}, pages 2024--10, 2024.

\bibitem{crampon2022machine}
Kevin Crampon, Alexis Giorkallos, Myrtille Deldossi, St{\'e}phanie Baud, and Luiz~Angelo Steffenel.
\newblock Machine-learning methods for ligand--protein molecular docking.
\newblock {\em Drug discovery today}, 27(1):151--164, 2022.

\bibitem{anderson2003process}
Amy~C Anderson.
\newblock The process of structure-based drug design.
\newblock {\em Chemistry \& biology}, 10(9):787--797, 2003.

\bibitem{achiam2023gpt}
Josh Achiam, Steven Adler, Sandhini Agarwal, Lama Ahmad, Ilge Akkaya, Florencia~Leoni Aleman, Diogo Almeida, Janko Altenschmidt, Sam Altman, Shyamal Anadkat, et~al.
\newblock Gpt-4 technical report.
\newblock {\em arXiv preprint arXiv:2303.08774}, 2023.

\bibitem{guo2025deepseek}
Daya Guo, Dejian Yang, Haowei Zhang, Junxiao Song, Ruoyu Zhang, Runxin Xu, Qihao Zhu, Shirong Ma, Peiyi Wang, Xiao Bi, et~al.
\newblock Deepseek-r1: Incentivizing reasoning capability in llms via reinforcement learning.
\newblock {\em arXiv preprint arXiv:2501.12948}, 2025.

\bibitem{liu2022swin}
Ze~Liu, Han Hu, Yutong Lin, Zhuliang Yao, Zhenda Xie, Yixuan Wei, Jia Ning, Yue Cao, Zheng Zhang, Li~Dong, et~al.
\newblock Swin transformer v2: Scaling up capacity and resolution.
\newblock In {\em Proceedings of the IEEE/CVF conference on computer vision and pattern recognition}, pages 12009--12019, 2022.

\bibitem{kirillov2023segment}
Alexander Kirillov, Eric Mintun, Nikhila Ravi, Hanzi Mao, Chloe Rolland, Laura Gustafson, Tete Xiao, Spencer Whitehead, Alexander~C Berg, Wan-Yen Lo, et~al.
\newblock Segment anything.
\newblock In {\em Proceedings of the IEEE/CVF international conference on computer vision}, pages 4015--4026, 2023.

\bibitem{taylor2022galactica}
Ross Taylor, Marcin Kardas, Guillem Cucurull, Thomas Scialom, Anthony Hartshorn, Elvis Saravia, Andrew Poulton, Viktor Kerkez, and Robert Stojnic.
\newblock Galactica: A large language model for science.
\newblock {\em arXiv preprint arXiv:2211.09085}, 2022.

\bibitem{fang2024mol}
Yin Fang, Xiaozhuan Liang, Ningyu Zhang, Kangwei Liu, Rui Huang, Zhuo Chen, Xiaohui Fan, and Huajun Chen.
\newblock Mol-instructions: A large-scale biomolecular instruction dataset for large language models.
\newblock In {\em The Twelfth International Conference on Learning Representations}.

\bibitem{luo2024biomedgpt}
Yizhen Luo, Jiahuan Zhang, Siqi Fan, Kai Yang, Massimo Hong, Yushuai Wu, Mu~Qiao, and Zaiqing Nie.
\newblock Biomedgpt: An open multimodal large language model for biomedicine.
\newblock {\em IEEE Journal of Biomedical and Health Informatics}, 2024.

\bibitem{pei2024biot5+}
Qizhi Pei, Lijun Wu, Kaiyuan Gao, Xiaozhuan Liang, Yin Fang, Jinhua Zhu, Shufang Xie, Tao Qin, and Rui Yan.
\newblock Biot5+: Towards generalized biological understanding with iupac integration and multi-task tuning.
\newblock In {\em Findings of the Association for Computational Linguistics ACL 2024}, pages 1216--1240, 2024.

\bibitem{vaswani2017attention}
Ashish Vaswani, Noam Shazeer, Niki Parmar, Jakob Uszkoreit, Llion Jones, Aidan~N Gomez, {\L}ukasz Kaiser, and Illia Polosukhin.
\newblock Attention is all you need.
\newblock {\em Advances in neural information processing systems}, 30, 2017.

\bibitem{hayes2025simulating}
Thomas Hayes, Roshan Rao, Halil Akin, Nicholas~J Sofroniew, Deniz Oktay, Zeming Lin, Robert Verkuil, Vincent~Q Tran, Jonathan Deaton, Marius Wiggert, et~al.
\newblock Simulating 500 million years of evolution with a language model.
\newblock {\em Science}, page eads0018, 2025.

\bibitem{wang2024dplm}
Xinyou Wang, Zaixiang Zheng, Fei Ye, Dongyu Xue, Shujian Huang, and Quanquan Gu.
\newblock Dplm-2: A multimodal diffusion protein language model.
\newblock {\em arXiv preprint arXiv:2410.13782}, 2024.

\bibitem{van2017neural}
Aaron Van Den~Oord, Oriol Vinyals, et~al.
\newblock Neural discrete representation learning.
\newblock {\em Advances in neural information processing systems}, 30, 2017.

\bibitem{sahoo2024simple}
Subham Sahoo, Marianne Arriola, Yair Schiff, Aaron Gokaslan, Edgar Marroquin, Justin Chiu, Alexander Rush, and Volodymyr Kuleshov.
\newblock Simple and effective masked diffusion language models.
\newblock {\em Advances in Neural Information Processing Systems}, 37:130136--130184, 2024.

\bibitem{xu2023geometric}
Minkai Xu, Alexander Powers, Ron Dror, Stefano Ermon, and Jure Leskovec.
\newblock Geometric latent diffusion models for 3d molecule generation.
\newblock In {\em International Conference on Machine Learning (ICML)}. NSF-PAR, 2023.

\bibitem{fu2024latent}
Cong Fu, Keqiang Yan, Limei Wang, Wing~Yee Au, Michael~Curtis McThrow, Tao Komikado, Koji Maruhashi, Kanji Uchino, Xiaoning Qian, and Shuiwang Ji.
\newblock A latent diffusion model for protein structure generation.
\newblock In {\em Learning on Graphs Conference}, pages 29--1. PMLR, 2024.

\bibitem{liproteinzen}
Alex~J Li and Tanja Kortemme.
\newblock Proteinzen: combining latent and se (3) flow matching for all-atom protein generation.

\bibitem{lipman2022flow}
Yaron Lipman, Ricky~TQ Chen, Heli Ben-Hamu, Maximilian Nickel, and Matthew Le.
\newblock Flow matching for generative modeling.
\newblock In {\em The Eleventh International Conference on Learning Representations}, 2022.

\bibitem{song2023unified}
Yuxuan Song, Jingjing Gong, Hao Zhou, Mingyue Zheng, Jingjing Liu, and Wei-Ying Ma.
\newblock Unified generative modeling of 3d molecules with bayesian flow networks.
\newblock In {\em The Twelfth International Conference on Learning Representations}, 2023.

\bibitem{nichol2021improved}
Alexander~Quinn Nichol and Prafulla Dhariwal.
\newblock Improved denoising diffusion probabilistic models.
\newblock In {\em International conference on machine learning}, pages 8162--8171. PMLR, 2021.

\bibitem{satorras2021n}
V{\i}ctor~Garcia Satorras, Emiel Hoogeboom, and Max Welling.
\newblock E (n) equivariant graph neural networks.
\newblock In {\em International conference on machine learning}, pages 9323--9332. PMLR, 2021.

\bibitem{molcraft}
Yanru Qu, Keyue Qiu, Yuxuan Song, Jingjing Gong, Jiawei Han, Mingyue Zheng, Hao Zhou, and Wei-Ying Ma.
\newblock Molcraft: Structure-based drug design in continuous parameter space.
\newblock In {\em International Conference on Machine Learning}, pages 41749--41768. PMLR, 2024.

\bibitem{xugeodiff}
Minkai Xu, Lantao Yu, Yang Song, Chence Shi, Stefano Ermon, and Jian Tang.
\newblock Geodiff: A geometric diffusion model for molecular conformation generation.
\newblock In {\em International Conference on Learning Representations}.

\bibitem{gebauer2019symmetry}
Niklas Gebauer, Michael Gastegger, and Kristof Sch{\"u}tt.
\newblock Symmetry-adapted generation of 3d point sets for the targeted discovery of molecules.
\newblock {\em Advances in neural information processing systems}, 32, 2019.

\bibitem{ichihara2011compound}
Osamu Ichihara, John Barker, Richard~J Law, and Mark Whittaker.
\newblock Compound design by fragment-linking.
\newblock {\em Molecular Informatics}, 30(4):298--306, 2011.

\bibitem{kong2022conditional}
Xiangzhe Kong, Wenbing Huang, and Yang Liu.
\newblock Conditional antibody design as 3d equivariant graph translation.
\newblock In {\em The Eleventh International Conference on Learning Representations}, 2022.

\bibitem{posebusters}
Martin Buttenschoen, Garrett~M Morris, and Charlotte~M Deane.
\newblock Posebusters: Ai-based docking methods fail to generate physically valid poses or generalise to novel sequences.
\newblock {\em Chemical Science}, 15(9):3130--3139, 2024.

\bibitem{diffdock}
Gabriele Corso, Hannes St{\"a}rk, Bowen Jing, Regina Barzilay, and Tommi Jaakkola.
\newblock Diffdock: Diffusion steps, twists, and turns for molecular docking.
\newblock {\em arXiv preprint arXiv:2210.01776}, 2022.

\bibitem{unimoldockingv2}
Eric Alcaide, Zhifeng Gao, Guolin Ke, Yaqi Li, Linfeng Zhang, Hang Zheng, and Gengmo Zhou.
\newblock Uni-mol docking v2: Towards realistic and accurate binding pose prediction.
\newblock {\em arXiv preprint arXiv:2405.11769}, 2024.

\bibitem{surfdock}
Duanhua Cao, Mingan Chen, Runze Zhang, Zhaokun Wang, Manlin Huang, Jie Yu, Xinyu Jiang, Zhehuan Fan, Wei Zhang, Hao Zhou, et~al.
\newblock Surfdock is a surface-informed diffusion generative model for reliable and accurate protein--ligand complex prediction.
\newblock {\em Nature Methods}, pages 1--13, 2024.

\bibitem{loshchilov2019decoupled}
Ilya Loshchilov and Frank Hutter.
\newblock Decoupled weight decay regularization.
\newblock In {\em International Conference on Learning Representations}, 2019.

\bibitem{pdbbind}
Zhihai Liu, Yan Li, Li~Han, Jie Li, Jie Liu, Zhixiong Zhao, Wei Nie, Yuchen Liu, and Renxiao Wang.
\newblock Pdb-wide collection of binding data: current status of the pdbbind database.
\newblock {\em Bioinformatics}, 31(3):405--412, 2015.

\bibitem{moad}
Mark~L Benson, Richard~D Smith, Nickolay~A Khazanov, Brandon Dimcheff, John Beaver, Peter Dresslar, Jason Nerothin, and Heather~A Carlson.
\newblock Binding moad, a high-quality protein--ligand database.
\newblock {\em Nucleic acids research}, 36(suppl\_1):D674--D678, 2007.

\bibitem{crossdocked}
Paul~G Francoeur, Tomohide Masuda, Jocelyn Sunseri, Andrew Jia, Richard~B Iovanisci, Ian Snyder, and David~R Koes.
\newblock Three-dimensional convolutional neural networks and a cross-docked data set for structure-based drug design.
\newblock {\em Journal of chemical information and modeling}, 60(9):4200--4215, 2020.

\bibitem{pepbdb}
Zeyu Wen, Jiahua He, Huanyu Tao, and Sheng-You Huang.
\newblock Pepbdb: a comprehensive structural database of biological peptide--protein interactions.
\newblock {\em Bioinformatics}, 35(1):175--177, 2019.

\bibitem{corso2024deep}
Gabriele Corso, Arthur Deng, Nicholas Polizzi, Regina Barzilay, and Tommi~S Jaakkola.
\newblock Deep confident steps to new pockets: Strategies for docking generalization.
\newblock In {\em The Twelfth International Conference on Learning Representations}.

\bibitem{trott2010autodock}
Oleg Trott and Arthur~J Olson.
\newblock Autodock vina: improving the speed and accuracy of docking with a new scoring function, efficient optimization, and multithreading.
\newblock {\em Journal of computational chemistry}, 31(2):455--461, 2010.

\bibitem{pocket2mol}
Xingang Peng, Shitong Luo, Jiaqi Guan, Qi~Xie, Jian Peng, and Jianzhu Ma.
\newblock Pocket2mol: Efficient molecular sampling based on 3d protein pockets.
\newblock In {\em International Conference on Machine Learning}, pages 17644--17655. PMLR, 2022.

\bibitem{targetdiff}
Jiaqi Guan, Wesley~Wei Qian, Xingang Peng, Yufeng Su, Jian Peng, and Jianzhu Ma.
\newblock 3d equivariant diffusion for target-aware molecule generation and affinity prediction.
\newblock {\em arXiv preprint arXiv:2303.03543}, 2023.

\bibitem{decompdiff}
Jiaqi Guan, Xiangxin Zhou, Yuwei Yang, Yu~Bao, Jian Peng, Jianzhu Ma, Qiang Liu, Liang Wang, and Quanquan Gu.
\newblock Decompdiff: diffusion models with decomposed priors for structure-based drug design.
\newblock {\em arXiv preprint arXiv:2403.07902}, 2024.

\bibitem{gao2024theory}
Bowen Gao, Haichuan Tan, Yanwen Huang, Minsi Ren, Xiao Huang, Wei-Ying Ma, Ya-Qin Zhang, and Yanyan Lan.
\newblock From theory to therapy: Reframing sbdd model evaluation via practical metrics.
\newblock {\em arXiv preprint arXiv:2406.08980}, 2024.

\bibitem{wu2025inference}
Yangzhen Wu, Zhiqing Sun, Shanda Li, Sean Welleck, and Yiming Yang.
\newblock Inference scaling laws: An empirical analysis of compute-optimal inference for llm problem-solving.
\newblock In {\em The Thirteenth International Conference on Learning Representations}, 2025.

\bibitem{shao2024deepseekmath}
Zhihong Shao, Peiyi Wang, Qihao Zhu, Runxin Xu, Junxiao Song, Xiao Bi, Haowei Zhang, Mingchuan Zhang, YK~Li, Y~Wu, et~al.
\newblock Deepseekmath: Pushing the limits of mathematical reasoning in open language models.
\newblock {\em arXiv preprint arXiv:2402.03300}, 2024.

\bibitem{yu2025dapo}
Qiying Yu, Zheng Zhang, Ruofei Zhu, Yufeng Yuan, Xiaochen Zuo, Yu~Yue, Tiantian Fan, Gaohong Liu, Lingjun Liu, Xin Liu, et~al.
\newblock Dapo: An open-source llm reinforcement learning system at scale.
\newblock {\em arXiv preprint arXiv:2503.14476}, 2025.

\end{thebibliography}

\appendix
\setcounter{section}{0}
\setcounter{equation}{0}
\setcounter{subsection}{0}
\setcounter{table}{0}
\setcounter{figure}{0}
\renewcommand{\theequation}{A\arabic{equation}}
\renewcommand{\thefigure}{A\arabic{figure}}
\renewcommand{\thetable}{A\arabic{table}}
\renewcommand{\thesubsection}{\Alph{subsection}}

\section*{Appendix}

\section{Detailed Model Architecture \label{app:arch}}
The network first combines noise information $\sigma$ with $\tilde{A}, \tilde{E}$ with embedding layers to extract dense features:
        
\begin{equation}
    \mathbf{h}=\text{Embed}(\tilde{A}\oplus \sigma_A),\quad \mathbf{e}=\text{Embed}(\tilde{E}\oplus \sigma_E)
\end{equation}

\paragraph{The protein and pocket channel} This channel calculates the representations of atoms within pockets and proteins by constructing a k-nearest neighbor graph $\mathcal{G}_{\text{poc}}$ by performing message passing and feature propagation with a 4-layer GNN. Each layer is composed of the following steps:
\begin{equation}
\begin{aligned}
    &\tilde{\mathbf{h}}^{(\text{poc})}_i\leftarrow \text{MLP}(\mathbf{h}^{(\text{poc})}_i),\\
    &m_{ij}\leftarrow \text{Linear}\left(\text{MLP}(\text{RBF}(d_{ij})+\mathbf{\tilde{h}}^{\text{(poc)}}_i+\mathbf{\tilde{h}}^{\text{(poc)}}_j\right),\\
    &\mathbf{h}^{(\text{poc})}_i\leftarrow \mathbf{h}^{(\text{poc})}_i+\text{LayerNorm}\left(\text{MLP}(\tilde{\mathbf{h}}^{(\text{poc})}_i)+\sum_{(i,j)\in \mathcal{G}_{\text{poc}}}m_{ij}\right),
\end{aligned}
\end{equation}

where $d_{ij}=\|\mathbf{x}_i-\mathbf{x}_j\|_2$ is the Euclidian distance between atom $i$ and $j$, $\text{MLP}$ is a multi-layer perceptron with two linear layers, ReLU activation and layer normalization, $\text{RBF}$ is the radial basis function using 32 Gaussian kernels ranging from 0 to 10, $\text{Linear}$ is a linear layer, and $\text{LayerNorm}$ denotes layer normalization.

\paragraph{The molecule channel} This channel updates the atomic and pairwise representations of atoms as well as their coordinates within small molecules that combine information from the protein and pocket channel. Similarly, a kNN graph $\mathcal{G}_{\text{poc-mol}}$ between pocket or protein atoms and molecule atoms are built. It also formulates a molecular graph $\mathcal{G}_{\text{mol}}$ based on $\tilde{E}$. Then, a 6-layer GNN is employed, each employing the following operation:

\begin{equation}
\begin{aligned}
&\mathbf{h}^{(\text{mol})}_i\leftarrow \mathbf{h}^{(\text{mol})}_i+\sum_{(i,j)\in \mathcal{G}_{\text{poc-mol}}}\phi_h\left(\mathbf{h}_i^{\text{(mol)}},\mathbf{h}^{(\text{poc})}_j,\mathbf{e}_{ij},d_{ij}\right)\\&\qquad \qquad \qquad \ \,+\sum_{(i,k)\in \mathcal{G}_{\text{mol}}}\phi_h\left(\mathbf{h}_i^{\text{(mol)}},\mathbf{h}^{(\text{mol})}_k,\mathbf{e}_{ik},d_{ik}\right),\\
&\mathbf{e}_{ij}^{(\text{mol})}\leftarrow \mathbf{e}_{ij}^{(\text{mol})}+\phi_e\left(\mathbf{h}_i^{\text{(mol)}},\mathbf{h}_j^{\text{(mol)}},\mathbf{e}_{ij}^{(\text{mol})}\right),\\
& \mathbf{x}_i^{(\text{mol})}\leftarrow \mathbf{x}_i^{(\text{mol})}+\sum_{(i,j)\in \mathcal{G}_{\text{poc-mol}}}(\mathbf{x}_j-\mathbf{x}_i)\phi_x\left(\mathbf{h}_i^{\text{(mol)}},\mathbf{h}_j^{\text{(poc)}},d_{ij}\right)\\&\qquad \qquad \qquad \ \,+\sum_{(i,k)\in \mathcal{G}_{\text{mol}}}(\mathbf{x}_k-\mathbf{x}_i)\phi_x\left(\mathbf{h}_i^{\text{(mol)}},\mathbf{h}_k^{\text{(mol)}},d_{ik}\right),
\end{aligned}
\end{equation}
where $\phi_h(\mathbf{h}_i,\mathbf{h}_j,d_{ij},\mathbf{e}_{ij})$ and $\phi_x(\mathbf{h}_i,\mathbf{h}_j,d_{ij},\mathbf{e}_{ij})$ are message-passing blocks that takes $\text{MLP}(\mathbf{h}_i)\oplus\text{MLP}(\mathbf{h}_j)\oplus\text{MLP}(\mathbf{e}_{ij})\oplus \text{MLP}(\text{RBF}(d_{ij}))$ as message ($\oplus$ denotes concatenation) to update $\mathbf{h}_i$, and $\phi_{e}(\mathbf{h}_i,\mathbf{h}_j,\mathbf{e}_{ij})$ updates $\mathbf{e}_{ij}$ with bi-directional messages by $\text{Linear}(\text{MLP}(\mathbf{h}_i)\oplus\text{MLP}(\mathbf{e}_{ij}))$ and $\text{Linear}(\text{MLP}(\mathbf{h}_j)\oplus\text{MLP}(\mathbf{e}_{ij}))$.

\paragraph{The denoising and confidence prediction heads} Finally, PharMolixFM directly takes $\mathbf{x}$ from the final layer as the denoised atom coordinates. The logits of the discrete atom and bond types are predicted by feeding the output representations $\mathbf{h}$ and $\mathbf{e}$ into two MLP classification heads. We also introduce 3 confidence heads composed of an MLP and softmax activation. The confidence scores for atom types and coordinates are calculated based on $\mathbf{h}$, and the confidence scores for bond types are calculated based on $\mathbf{e}$.

\end{document}